\documentclass[prb,twoside,twocolumn,showpacs,superscriptaddress]{revtex4}
\usepackage{graphicx}
\usepackage{amsmath}
\usepackage{amssymb}

\begin{document}

\title{Disorder-induced Phase Transition of Vortex Matter in MgB$_2$}
\author{M. Angst}
 \email[Email: ]{angst@phys.ethz.ch}
\affiliation{Solid State Physics Laboratory ETH, 8093 Z\"urich,
Switzerland}
\author{R. Puzniak}
\author{A. Wisniewski}
\affiliation{Institute of Physics, Polish Academy of Sciences,
Aleja Lotnikow 32/46, 02-668 Warsaw, Poland}
\author{J. Jun}
\author{S. M. Kazakov}
\author{J. Karpinski}
\affiliation{Solid State Physics Laboratory ETH, 8093 Z\"urich,
Switzerland}
\date{\today}
\begin{abstract}
Measurements of single crystalline MgB$_2$ with torque
magnetometry in fields up to $90\,{\text{kOe}}$ reveal a sharp
peak in the irreversible torque at about $0.85\,H_{c2}$. In the
region between peak onset and maximum, pronounced history effects
occur. Angle and temperature dependence of the characteristic peak
fields track those of $H_{c2}$. The features observed suggest that
the peak marks a disorder-induced phase transition of vortex
matter between a quasi-ordered Bragg glass and a highly-disordered
phase.
\end{abstract}
 \maketitle


The new superconductor MgB$_2$ is considered to have great
potential for applications, and a lot of research activity has
concentrated on this compound.\cite{note_Buzea} Much light was
shed on the superconducting mechanism, there is mounting evidence
that MgB$_2$ is a two-band superconductor with a substantial
difference between the superconducting gaps of the two
bands.\cite{note_Buzea,Bouquet01c,Schmidt02,Iavarone02} About the
superconducting phase diagram, however, less is known. Most
``phase diagrams" published contain only the upper critical fields
$H_{c2}$. Here, the situation was clarified insofar as later bulk
measurements all find a pronounced anisotropy $\gamma$ of
$H_{\text{c2}}$, decreasing with increasing
temperature,\cite{Angst02MgB2anis,Sologubenko01,Budko02,Zehetmayer02,Welp02}
although there are still discrepancies of the exact $\gamma(T)$
dependences reported. MgB$_2$ is, particularly concerning the
importance of thermal fluctuations and the value of
$\kappa=\lambda/\xi$, intermediate between the high $T_c$ cuprates
and low $T_c$ superconductors. Studying the vortex matter phase
diagram of MgB$_2$ may thus help in understanding the phase
diagrams of various superconductors in a unified way.

From the study of cuprate superconductors is known that the
$H$$-$$T$ phase diagram contains more transition lines than the
upper and lower critical fields. Identified were for example a
melting transition between a quasi-ordered vortex lattice, called
Bragg glass, and a disordered vortex fluid,\cite{Schilling96} as
well as an order-disorder transition between the Bragg glass and a
highly disordered, glassy
phase.\cite{Giamarchi94,Khaykovich96,Giller97} The latter
transition\cite{note_Avraham} has been observed also in low $T_c$
superconductors, such as NbSe$_2$,\cite{Ravikumar01,Marchevsky01}
and even in the elemental superconductor Nb,\cite{Ling01} but not
in ultra-pure Nb crystals.\cite{Forgan02} This transition is
generally associated with a peak in the critical current density
$j_c$ and pronounced history effects.

In single crystals of MgB$_2$, a quasi-ordered vortex structure
has been observed in low fields by scanning tunneling
spectroscopy,\cite{Eskildsen02} showing that at least under some
conditions a Bragg glass is the stable vortex phase. Since, by
tuning the amount of quenched random point-like disorder, the
stabilization of a highly disordered phase can always be favored,
an order-disorder transition in fields below $H_{c2}$ should be
observable in MgB$_2$ as well, at least for certain impurity
concentrations. Although a phase transition distinct from $H_{c1}$
or $H_{c2}$ has not been suggested yet in MgB$_2$, a peak effect,
and accompanying history effects have been observed in transport
measurements for $H\|c$.\cite{Welp02}

Here, we report the observation of a pronounced, sharp peak effect
(PE) by torque magnetometry in fields close to, but clearly
distinct from, $H_{c2}$. A minor hysteresis loop (MHL) study shows
pronounced history effects in the region between the onset and the
maximum of the peak. Angle and temperature dependence of the
characteristic fields is reported, and we propose a phase diagram
for MgB$_2$.

\begin{figure}[tb]
\includegraphics[width=0.95\linewidth]{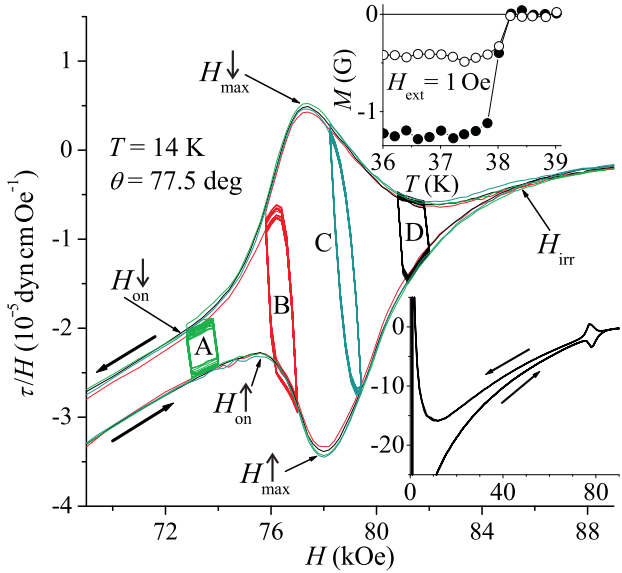}
\caption{Torque $\tau/H$ vs field $H$ at $14\,{\text{K}}$ and
$77.5\,{\text{deg}}$. The direction of the field change is
indicated by thick arrows. The irreversibility field
$H_{\text{irr}}$ and the onset and maximum fields $H_{\text{on}}$
and $H_{\text{max}}$ of the PE for the $H$ increasing
($^\uparrow$) and decreasing ($^\downarrow$) branch are marked.
Also shown are some of the MHL (see text) measured, labeled A-D.
Upper inset: $M(T)$ curve in the transition region, in a field
$H\|c$ of $1\,{\text{Oe}}$, zero field cooled ($\bullet$) and
field cooled ($\circ$). Lower inset: $\tau/H$ vs $H$ of the curve
in the main panel, for the whole field range.} \label{Fig1}
\end{figure}

The measurements were performed on a high-quality single crystal
of MgB$_2$, sample B of Ref.\ \onlinecite{Angst02MgB2anis}. The
$T$ dependence of the magnetization (upper inset of Fig.\
\ref{Fig1}) shows a sharp ($0.3\,{\text{K}}$ with a $10\%-90\%$
criterion) transition to the superconducting state at $38.2\,
{\text{K}}$, indicating a high quality of the crystal.
Measurements to study the PE were carried out with the torque
option of a Quantum Design 9T PPMS. Measurement runs consisted in
varying the field $H$ at fixed angle $\theta$ between
$\overrightarrow{H}$ and the $c$-axis of the crystal, and
recording the torque
$\overrightarrow{\tau}=\overrightarrow{m}\times\overrightarrow{H}$,
where $\overrightarrow{m}$ is the magnetic moment of the crystal.


%

One of the curves measured is shown in the lower inset of Fig.\
\ref{Fig1}. For better comparison with magnetization curves,
$\tau/H$ vs $H$ is shown. The main panel shows a magnification of
the PE region. The peak is well pronounced and very sharp. Various
characteristic fields are indicated: The maximum of the peak for
field increasing ($H_{\text{max}}^{\uparrow}$) and decreasing
($H_{\text{max}}^{\downarrow}$) branch of the hysteresis loop, and
the onsets of the peak, $H_{\text{on}}^{\uparrow}$ and
$H_{\text{on}}^{\downarrow}$. The separation of the two onset
fields is larger, similar to the case of the cuprate
superconductors (see, e.g., Ref.\ \onlinecite{Angst02Sr124}). Also
indicated is the irreversibility field $H_{\text{irr}}$, where the
two branches of the hysteresis loops meet. The peak resembles
qualitatively peaks observed in
NbSe$_2$,\cite{Ravikumar01,Eremenko02} and CeRu$_2$.\cite{Roy00}

To investigate possible history dependences of $j_c$, we performed
several minor hysteresis loop (MHL) measurements in and around the
peak: The field is cycled up and down by a small amount several
times, ideally until the loops retrace each other, indicating that
the vortex system reached a stable pinned state in the given
field.\cite{Ravikumar01,Roy00} MHL measured, within full loops, in
four different regions of the PE are indicated in the figure
(A-D).

\begin{figure}[tb]
\includegraphics[width=0.95\linewidth]{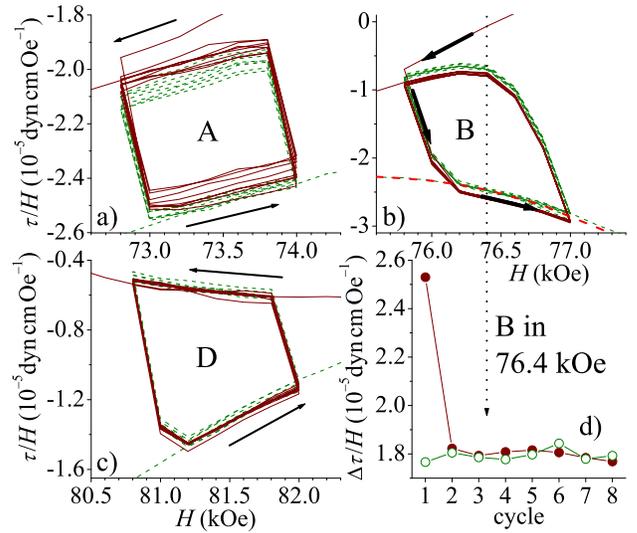}
\caption{a),b),c) Magnification of MHL A, B, and D also displayed
in Fig.\ \ref{Fig1}. MHL started from the field increasing branch
of the full hysteresis loop are shown as dotted lines, while those
started from the field decreasing branch are shown as full lines.
d) Width of the hysteresis of MHL B started from the field
increasing/decreasing ($\circ$/$\bullet$) branch of the full
hysteresis loop, as a function of cycling.} \label{Fig2}
\end{figure}

Torque $\tau/H$ values of MHL A (Fig.\ \ref{Fig2}a)) vary
significantly as the MHL is cycled through repeatedly. Partly,
this may be explained by relatively strong normal relaxation
processes. However, a pronounced difference can be seen between
MHL started from the field increasing ($H^{\uparrow}$) branch of
the full hysteresis loop (FHL), and the one started from the field
decreasing ($H^{\downarrow}$) branch. The latter has a
significantly higher width initially. This effect can be explained
by a difference in the vortex configuration between $H^{\uparrow}$
and $H^{\downarrow}$ in the region of MHL A. In the configuration
on $H^{\downarrow}$, $j_{c}$ (proportional to the width of the
MHL)\cite{note_rev_critst} is higher, i.e.\ the vortices are
pinned stronger. Repeated cycling causes the width of the MHL
started from $H^{\downarrow}$ to approach the one started from
$H^{\uparrow}$, indicating that the vortex configuration on
$H^{\downarrow}$ is only meta-stable. History effects are even
more pronounced for MHL B (Fig.\ \ref{Fig2}b)). Here, the initial
$H^{\uparrow}$ branch of the MHL started from the $H^{\downarrow}$
branch of the FHL (full line indicated by arrows) clearly is below
the $H^{\uparrow}$ branch of the FHL (thick dashed), indicating
larger hysteresis. This behavior contradicts Bean's critical state
model,\cite{Bean62} where the hysteresis of partial hysteresis
loops can never be higher than the one of the full loop. We point
out that simple relaxation effects cannot account for this
specific effect. It can be explained by the vortex configuration
on the $H^{\downarrow}$ branch of the FHL (where the MHL was
started) having a higher $j_c$ than the vortex configuration on
the $H^{\uparrow}$ branch. The variation of the hysteresis width
with cycling (Fig.\ \ref{Fig2}d)) demonstrates the meta-stable
nature of the vortex configuration on the $H^{\downarrow}$ branch
of the FHL, while the vortex configuration of the $H^{\uparrow}$
branch of the FHL is stable, or close to. In contrast, no clear
deviations in the cycling behavior between $H^{\uparrow}$ and
$H^{\downarrow}$ branch started MHL are visible for MHL C and MHL
D (Fig.\ \ref{Fig2}c)), as well as for a MHL measured in the
region around $68\,{\text{kOe}}$ (not shown).

In summary, between $H_{\text{on}}^{\downarrow}$ and
$H_{\text{max}}^{\uparrow}$, pronounced history effects occur.
They can be accounted for by the coexistence of a meta-stable
high-field vortex configuration with high pinning and a stable
low-field, low pinning configuration. Above
$H_{\text{max}}^{\uparrow}$ and below
$H_{\text{on}}^{\downarrow}$, no significant history effects are
observed, indicating that there is only one vortex configuration,
which is stable. The larger hysteresis width of MHL started from
$H^{\downarrow}$ indicates pinning in the configuration stable
above $H_{\text{max}}^{\uparrow}$ to be stronger than pinning in
the configuration stable below $H_{\text{on}}^{\downarrow}$.

\begin{figure}[tb]
\includegraphics[width=0.95\linewidth]{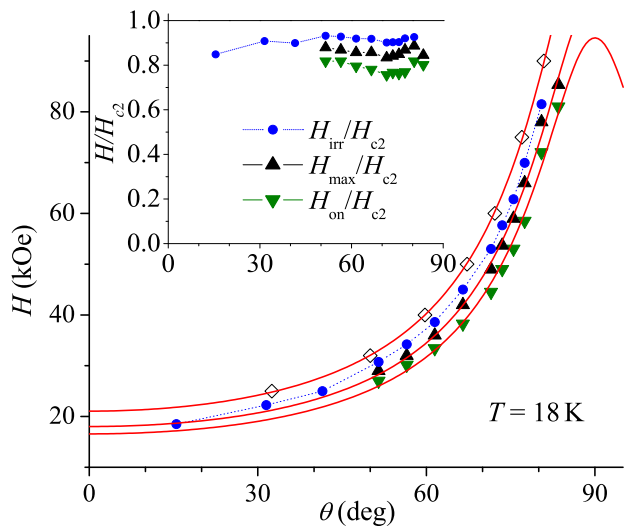}
\caption{Angle dependence of various characteristic fields at
$18\,{\text{K}}$. Shown are the upper critical field
$H_{\text{c2}}$ ($\diamondsuit$, from Ref.\
\onlinecite{Angst02MgB2anis}), the irreversibility field
$H_{\text{irr}}$ ($\bullet$),\cite{note_Hirr_crit} the peak
maximum field $H_{\max}$ ($\blacktriangle$), and the peak onset
field $H_{\text{on}}$ ($\blacktriangledown$). Full lines are fits
of the theoretical $H_{c2}(\theta)$
dependence.\cite{Angst02MgB2anis} Dashed lines are guides for the
eye. Inset: Angle dependence of reduced (divided by
$H_{\text{c2}}(\theta)$) characteristic fields.} \label{Fig3}
\end{figure}

The variation of the peak onsets and maxima with angle at
$18\,{\text{K}}$ is shown, together with $H_{c2}(\theta)$ and
$H_{\text{irr}}(\theta)$,\cite{note_Hirr_crit} in Fig.\
\ref{Fig3}. Since the visibility of the peaks is diminished at
higher temperatures, onsets and maxima were determined from
$\Delta \tau (H) = \tau(H^{\downarrow})-\tau(H^{\uparrow})$
curves. The characteristic peak fields follow the angular
dependence of $H_{c2}$, as indicated by fits to the theoretical
$H_{c2}(\theta)$ dependence according to the anisotropic
Ginzburg-Landau theory (see Ref.\ \onlinecite{Angst02MgB2anis}),
while the angular scaling of the irreversibility field is less
clear. This can be seen also in the inset, displaying the $\theta$
dependence of the characteristic fields, reduced by the upper
critical field. The onset field is approximately constant at about
$0.8\,H_{c2}$ and the maximum field at about $0.85\,H_{c2}$.
$H_{\text{irr}}$ is located at about $0.9\,H_{c2}$, but seems to
get slightly lower as $\theta\rightarrow 0$.

The characteristic fields could not be determined with enough
accuracy in the whole region of angles: Since
$\overrightarrow{\tau}=\overrightarrow{m}\times\overrightarrow{H}$
and $\overrightarrow{m}$ points, for $H\|c$ or $H\|ab$, into the
same direction, the sensitivity is much lower for angles close to
$0$ and $90\,{\text{deg}}$. Due to the pronounced anisotropy of
MgB$_2$ at $18\,{\text{K}}$ ($\gamma\simeq
5.7$)\cite{Angst02MgB2anis} $\overrightarrow{m}$ tends to be
directed almost perpendicular to the planes, except at very high
angles. Therefore, the maximum effective sensitivity of the torque
magnetometer is achieved at angles in the region of
$75$-$80\,{\text{deg}}$. SQUID measurements performed on the same
crystal with $H\|c$ and $H\|ab$ showed no sign of a PE in the
region around $0.8\,H_{c2}$. This is likely due to insufficient
sensitivity of the SQUID and field inhomogeneities in the SQUID
magnetometer, which, due to the movement of the sample, tend to
smear such features.\cite{Ravikumar97} However, preliminary ac
susceptibility data\cite{Puzniak_inprep} measured on the same
crystal indicate the PE to be present both for $H\|c$ and $H\|ab$,
confirming that the underlying mechanism is a feature for all
directions of $H$.

\begin{figure}[tb]
\includegraphics[width=0.95\linewidth]{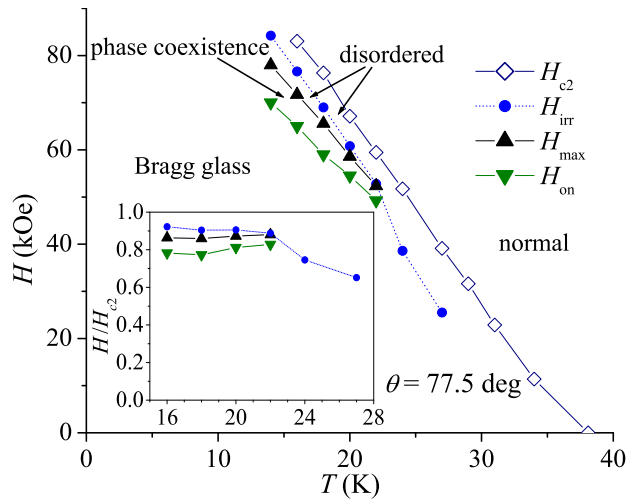}
\caption{Phase diagram of MgB$_2$ single crystal at an angle of
$77.5\,{\text{deg}}$ between the $c$-axis of the crystal and the
applied field: The temperature dependence of the characteristic
fields $H_{\text{c2}}$, $H_{\text{max}}$ and $H_{\text{on}}$ is
given. They mark boundaries between the normal state and the
various phases of vortex matter. The irreversibility field
$H_{\text{irr}}$ is also shown. The inset shows the $T$ dependence
of the characteristic fields scaled by $H_{\text{c2}}$.}
\label{Fig4}
\end{figure}

In Fig.\ \ref{Fig4}, the $T$ dependence of the characteristic
fields is shown, for $77.5\,{\text{deg}}$, which corresponds
roughly to the angle where the PE is most visible. The peak
amplitude is reduced quickly by increasing $T$, and above
$22\,{\text{K}}$ the PE is no longer clearly discernible in the
$\tau(H)$ data. This is due to the decreased sensitivity of the
magnetometer\cite{note_sensH} and due to thermal smearing of the
effective disorder potential. The inset shows that the positions
of $H_{\text{on}}$ and $H_{\text{max}}$ relative to $H_{c2}$ are
approximately constant. It is, therefore, unlikely that they would
merge with the upper critical field at some higher temperature. In
contrast, $H_{\text{irr}}$ shifts to lower fields relative to
$H_{c2}$ as $T$ increases, also likely due to a smearing of the
effective pinning landscape by thermal fluctuations. There is no
indication that $H_{\text{irr}}$ may correspond to a phase
transition.

Before discussing the PE in terms of a disorder-induced phase
transition of vortex matter, we briefly examine alternative
origins of the PE. The possibility that the PE is due to
inhomogeneities or extended defects is not likely. A second
crystal grown with the same technique, but under slightly
different conditions, shows also a PE, in similar fields. The
pinning properties of the two crystals for $H$ nearly aligned
parallel to the $ab$ planes are different in a pronounced
way.\cite{CommentTaka} It therefore seems rather unlikely that the
two crystals would have identical structural, non-intrinsic
features leading to a similar PE. A further possibility would be a
change of the elastic constants of the vortex lattice when $H$
approaches $H_{c2}$, not associated with a phase
transition.\cite{Pippard69} However, the specific form of the
history effects observed in the PE region are hard to explain
without a phase transition. Thermal melting can rather be
excluded. Thermal fluctuations should be much less important in
MgB$_2$ than in the cuprate superconductors: MgB$_2$ has a
Ginzburg number $Gi=\frac{1}{2}(\gamma k_B T_c
/H_c^2(0)\xi_{ab}^3(0))^2$, a measure of the importance of thermal
fluctuations, of the order of $10^{-5}$ only, while the cuprates
typically have $Gi\approx 10^{-2}$.\cite{Mikitik01} On the other
hand, thermal fluctuations should be more important than for
example in Nb with $Gi\approx10^{-10}$,\cite{Forgan02} or NbSe$_2$
with $Gi\approx10^{-8}$.\cite{Eremenko02} A calculation of the
melting field $H_m$, using Eq.\ (26) of Ref.\
\onlinecite{Mikitik01} leads, at $14\,{\text{K}}$, to
$H_m/H_{c2}\approx 0.97$, much higher than the location of the
peak and therefore hardly can account for it,\cite{note_PEmelt}
although it was shown that point disorder can shift $H_m$ to
slightly lower fields.\cite{Nishizaki00} Also, a liquid caused by
thermal melting should have weaker pinning properties than the
solid lattice.

An important fact deduced from the MHL experiments is that the
high field phase has got a higher critical current density than
the low field phase. This is the case for the transition from a
Bragg glass to a highly disordered glassy phase.\cite{Giamarchi94}
That we indeed observed this phase transition is supported by the
pronounced history dependence of $j_c$ in the region between onset
and maximum of the peak, of a form similar to observations of the
PE in NbSe$_2$ and not accountable for by relaxation effects. The
location relatively close to $H_{c2}$ is expected for a
superconductor with low $Gi$ and relatively weak
disorder.\cite{Mikitik01} In NbSe$_2$, a superconductor with
comparable, but even lower $Gi$, there is conclusive evidence that
the PE is indeed due to the transition between a Bragg glass and a
highly disordered phase.\cite{Marchevsky01} The history effects
mark the region of meta-stability, where a macroscopic coexistence
of the two phases is possible. Pinning of the phase boundary is
directly responsible for the history effects. The location of the
PE with respect to $H_{c2}$, together with the history effects
studied and the observation of a higher critical current density
in the high field vortex configuration, thus indicate that the PE
in MgB$_2$ marks the transition between the Bragg glass and a
highly disordered phase, which may be termed ``amorphous'' or
``pinned liquid''. If the PE observed by Welp {\em{et
al.}}\cite{Welp02} is of the same origin, the larger separation of
the PE from $H_{c2}$ in our case indicates that the crystal
investigated by us has a higher amount of random point-like
disorder. Further investigations of the transition line with
controlled tuning of the amount of disorder, as was done in the
case of the cuprates, by electron irradiation\cite{Nishizaki00}
and chemical substitution,\cite{Angst02Sr124} may help finding a
unified description of the phase diagrams of different
superconductors.

In summary, using torque magnetometry, we observed a pronounced,
sharp peak effect in single crystalline MgB$_2$. Onset and maximum
of the peak are located at about $0.8\,H_{c2}$ and $0.85\,H_{c2}$,
with little dependence on the temperature or the direction of the
applied field. Peak form, history effects between onset and
maximum, as well as the location of the peak are consistent with
the peak effect marking a phase transition between the Bragg glass
and a highly disordered phase of vortex matter.

\label{ack} We thank J.~Roos for useful discussions. This work was
supported by the Swiss National Science Foundation, by the
European Community (contract ICA1-CT-2000-70018) and by the Polish
State Committee for Scientific Research (5 P03B 12421).

{\em Note} After submission of this manuscript, we became aware of
a report\cite{Pissas02} on ac susceptibility measurements for
$H\|c$, coming to similar conclusions.

\newcommand{\noopsort}[1]{} \newcommand{\printfirst}[2]{#1}
  \newcommand{\singleletter}[1]{#1} \newcommand{\switchargs}[2]{#2#1}


\begin{thebibliography}{10}
\bibitem{note_rev_critst}
The form of the MHL clearly indicates that a full reversal of the
critical state \cite{Bean62} is achieved.
\bibitem{note_Hirr_crit}
A criterion $\tau_{\text{irr}}/H\sin(2\theta)=10^{-6}
  {\text{dyn}}\,{\text{cm}}\,{\text{Oe}}^{-1}$ was used for defining
  $H_{\text{irr}}$.
\bibitem{note_sensH}
The signal-to-noise ratio is decreased in lower fields, since
$\tau \propto H$.
\end{thebibliography}
\end{document}